\documentclass[twocolumn,amsmath,amssymb,prx,aps,superscriptaddress]{revtex4-1}
\usepackage[version=3]{mhchem} 
\usepackage{amsmath}
\usepackage{upgreek}
\usepackage{color}
\usepackage{natbib}
\usepackage{graphicx,subfigure}
\usepackage{braket}
\usepackage{siunitx}

\usepackage{xspace}
\usepackage{textcomp}

\begin{document}

\title{Dynamic Force Measurements on Swimming \textit{Chlamydomonas} Cells using Micropipette Force Sensors}

\author{Thomas J. B\"{o}ddeker}\email{Present address: ETH Z\"urich, Switzerland.}
\affiliation{Max Planck Institute for Dynamics and Self-Organization (MPIDS), Am Fa{\ss}berg 17, D-37077 G\"{o}ttingen, Germany}

\author{Stefan Karpitschka}
\affiliation{Max Planck Institute for Dynamics and Self-Organization (MPIDS), Am Fa{\ss}berg 17, D-37077 G\"{o}ttingen, Germany}

\author{Christian T. Kreis}
\affiliation{Max Planck Institute for Dynamics and Self-Organization (MPIDS), Am Fa{\ss}berg 17, D-37077 G\"{o}ttingen, Germany}

\author{Quentin Magdelaine}
\affiliation{Max Planck Institute for Dynamics and Self-Organization (MPIDS), Am Fa{\ss}berg 17, D-37077 G\"{o}ttingen, Germany}

\author{Oliver B\"{a}umchen} \email{Corresponding author: oliver.baeumchen@ds.mpg.de.}
\affiliation{Max Planck Institute for Dynamics and Self-Organization (MPIDS), Am Fa{\ss}berg 17, D-37077 G\"{o}ttingen, Germany}

\date{\today}

\begin{abstract} 
Flagella and cilia are cellular appendages that inherit essential functions of microbial life including sensing and navigating the environment. 
In order to propel a swimming microorganism they displace the surrounding fluid by means of periodic motions, while precisely-timed modulations of their beating patterns enable the cell to steer towards or away from specific locations.
Characterizing the dynamic forces, however, is challenging and typically relies on indirect experimental approaches.
Here, we present direct \textit{in vivo} measurements of the dynamic forces of motile \textit{Chlamydomonas reinhardtii} cells in controlled environments. 
The experiments are based on partially aspirating a living microorganism at the tip of a micropipette force sensor and optically recording the micropipette's position fluctuations with high temporal and sub-pixel spatial resolution. 
Spectral signal analysis allows for isolating the cell-generated dynamic forces associated to the periodic motion of the flagella from background noise. 
We provide an analytic elasto-hydrodynamic model for the micropipette force sensor and describe how to obtain the micropipette's full frequency response function from a dynamic force calibration. 
Using this approach, we find dynamic forces during the free swimming activity of individual \textit{Chlamydomonas reinhardtii} cells of $23\pm 5$\,pN resulting from the coordinated flagellar beating with a frequency of $51\pm 6$\,Hz. 
This dynamic micropipette force sensor (DMFS) technique generalises the applicability of micropipettes as force sensors from static to dynamic force measurements, yielding a force sensitivity in the piconewton range.
In addition to measurements in bulk liquid environment, we study the dynamic forces of the biflagellated microswimmer in the vicinity of a solid/liquid interface. 
As we gradually decrease the distance of the swimming microbe to the interface, we measure a significantly enhanced force transduction at distances larger than the maximum extend of the beating flagella, highlighting the importance of hydrodynamic interactions for scenarios in which flagellated microorganisms encounter surfaces.
\end{abstract}

\maketitle


\begin{figure*}[t]
\centering
	\includegraphics[width=1\textwidth]{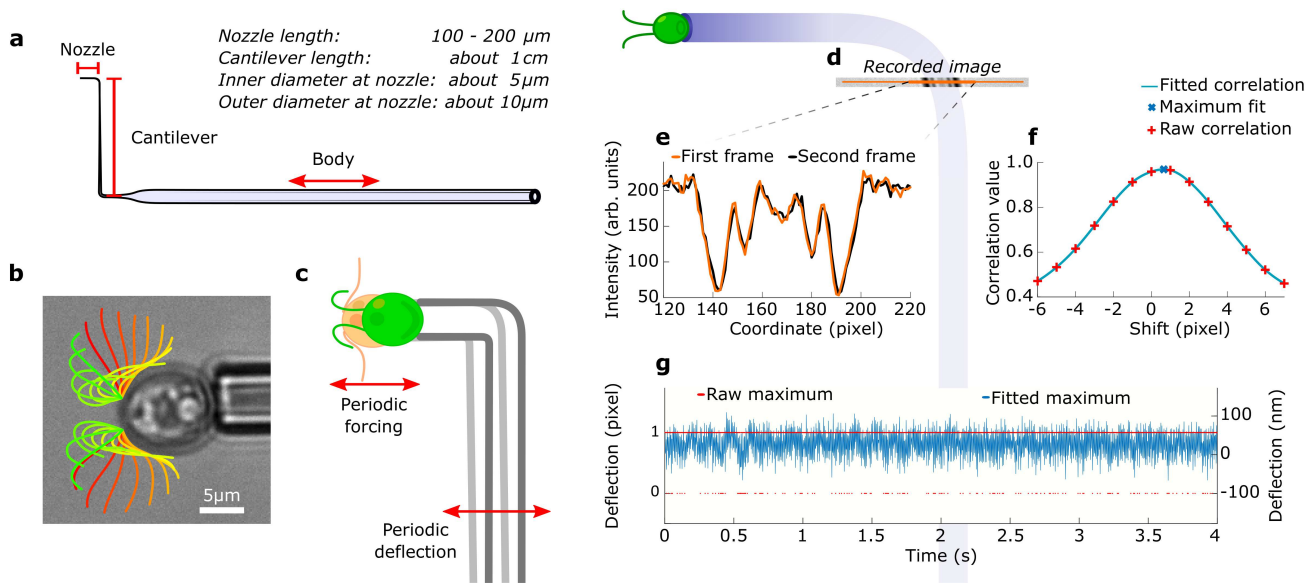}
\caption{
Dynamic force measurements on swimming \textit{C.~reinhardtii} cells using micropipette force sensors. 
\textbf{a} Sketch of the micropipette force sensor.
\textbf{b} Micrograph of a freely swimming \textit{C.~reinhardtii} cell fixated at the tip of the nozzle overlaid with the flagella shapes throughout one beating cycle. Time is color coded from red to green. The time step between snapshots is 1.55\,ms, corresponding to a flagella beating frequency of about 46 Hz. The original movie is provided as \textbf{Supplementary Movie}.
\textbf{c} The cell is held at the nozzle of the micropipette and exerts a periodic forcing that leads to a periodic deflection of the cantilever.
\textbf{d} Exemplar micrograph of the pipette cross-section recorded at the end of the cantilever. 
\textbf{e} Intensity profiles of two subsequent cross-section images.
\textbf{f} Correlation analysis of two frames to determine the relative shift corresponding to the maximal correlation. Interpolating the discrete correlation function results in sub-pixel deflection resolution. 
\textbf{g} Discrete and sub-pixel micropipette deflection over time. }

\label{Fig1}
\end{figure*}

\section{Introduction}
Microbial motility governs a variety of phenomena in microbiology, biophysics, as well as many other fields of research, and has been studied intensively for decades \cite{purcell,berg}. 
The propulsion of bacteria, microalgae, nematodes and other single- and multicellular microorganisms in a liquid medium can be either realized by the periodic motion of cellular appendages such as flagella and cilia, as for single-flagellated spermatozoa and the multiflagellated model microbes \textit{E.~coli} and \textit{C.~reinhardtii}, or by oscillatory shape changes, e.g.\ in case of the model nematode \textit{C.~elegans} \cite{Lauga2009}.
Microbes exhibit rich dynamics in their motility patterns.
The perception of gradients of, for example, chemicals, light intensity or fluid flow, may cause precisely-timed modulations of the periodic motions of cellular appendages.
In consequence, microbes can navigate towards regions of high nutrient concentration (chemotaxis) \cite{Berg1972,Alon1999}, optimal light intensity (phototaxis) \cite{Foster1984} or align relative to fluid flow (rheotaxis) \cite{Miki2013}. 
The study of microbial motility in general, and the response of motile microbes to environmental cues in particular, calls for experimental methods allowing to characterize the magnitude of the periodic forces underlying microbial propulsion.
However, directly measuring the dynamic forces generated by cellular appendages of motile microbes is challenging due to the size of the organisms and the magnitude of the forces, which are often just a few microns and piconewton, respectively.
Moreover, measuring the forces generated by motile cells in their planktonic state, i.e.\ freely swimming in a liquid medium, demands for experimental techniques that do not rely on supporting the cell by a substrate as in conventional atomic force microscopy approaches \cite{alsteens2017,krieg2019}.
Ultimately, one wishes to quantify also the modulations of these forces due to environmental cues, which requires simultaneous access to chemical composition, light intensity, fluid flow or other relevant parameters in the vicinity of the swimming microorganism.

To date, experimental techniques for characterizing the swimming motility of microbes largely rely on optical measurements and on the trapping of self-propelled cells by optical tweezers \cite{ashkin1987,min2009,stellamanns2014}. 
\textit{Chlamydomonas reinhardtii}, a unicellular, photoactive microalga with a cell diameter of about 8 to 10\,\textmu{}m featuring two anterior flagella, is a prime model organism in microbial motility studies.
Measurements of its swimming dynamics have been carried out by tracking the cell body \cite{Rueffer1985,Minoura1995,Polin2009,Guasto2010,Ostapenko2018}, particle imaging velocimetry (PIV) of the fluid around the organism \cite{Guasto2010,Drescher2010,Brumley2014}, and by analyzing the spatiotemporal changes of the flagella shape during its swimming strokes \cite{Rueffer1985,Rueffer1987,Polin2009,Goldstein2009,Bayly2011,Brumley2014,Wan2014,Wan2014b}.
The main drawback of these methods is their indirect approach, requiring a model for the flagellar hydrodynamics, typically using slender-body and resistive force theory \cite{Johnson1979}, to obtain estimates of the acting forces.
Direct force measurements on \textit{C.~reinhardtii} have so far only been realized using optical tweezers \cite{McCord2005}. 
This approach, however, is limited to the measurement of the maximum (escape) force per cell and cannot provide time-resolved force measurements.
Since biophysical techniques enabling truly direct force measurements are lacking to date, a conclusive picture of the mechanics and coordination of cilia and flagella, which control microbial motility, still remains elusive \cite{Wan2018}.
This includes an ongoing debate whether hydrodynamic or steric contact forces dominate the interaction of puller-type microswimmers with interfaces, as well as their impact on microbial navigation in confined spaces \cite{Kantsler2013,Contino2015}.

Attaching individual microorganisms to hollow glass micropipettes by applying a negative pressure on the inside of the pipette is a common technique to keep motile cells in the focal plane of the microscope. 
It allows to track the dynamics of their freely beating cilia or flagella \cite{Rueffer1987,Polin2009,Goldstein2009,Bayly2011,Brumley2014,Wan2014,Wan2014b}. 
In principle, forces can be measured simultaneously through the deflection of the micropipette.
However, previous works utilizing such micropipette force sensors \cite{Colbert2009,Colbert2010,Backholm2013,kreis2018,petit2018,backholm2019,kreis2019} were limited to quasi-static measurements.
In this case the damping of the micropipette cantilever by hydrodynamic drag forces remains negligible, and forces can be directly recovered from the cantilever deflection and its static spring constant.
In the context of microbial propulsion, such micropipette force sensors have been applied successfully to the millimetre-sized nematode \textit{C.~elegans} \cite{Schulman2014,Schulman2014b,Backholm2015}, which has a typical beating frequency of about 2\,Hz only \cite{Schulman2014}.
Many flagellated and ciliated micro-swimmers, however, operate at much higher frequencies.
In addition, they generate forces that are smaller by about two to three orders of magnitude, compared to the reported nanonewton propulsion forces of \textit{C.~elegans}.
Time-resolved measurements of such forces have been impossible so far because hydrodynamic drag forces become significant in the frequency range of the cantilever actuation.

Here, we present the first direct measurements of the dynamic forces originating from the periodic flagella beating of a living microorganism.
A highly flexible, double-L-shaped micropipette serves as a calibrated dynamic force cantilever.
An individual {\it C. reinhardtii} cell is attached to the tip of the micropipette such that forces exerted in the direction of its regular swimming motion are orthogonal to the cantilever arm, see \textbf{Fig.~\ref{Fig1}a} to \textbf{c}.
The pipette deflection is recorded optically with high spatial and temporal resolution.
We establish a versatile method for a full calibration of the micropipette's frequency response function and show that signal analysis in Fourier space allows for isolating the signal caused by the beating flagella of \textit{C.~reinhardtii} from external vibrations and background noise fluctuations.
This novel experimental approach enables direct measurement of dynamic forces in the piconewton range, generated by flagellated cells at high frequencies. Simultaneously, important environmental control parameters can be accessed.
We resolve the forces generated by the beating flagella of {\it C.~reinhardtii} as a function of their distance to a solid interface, elucidating the nature of wall interaction mechanisms of puller-type microswimmers.

\section{Signatures of Cell-Generated Dynamic Forces}
\subsection{Dynamic Micropipette Force Sensors}

The micropipette force sensor used to measure the dynamic forces exerted by the cell's flagella is custom-made, largely following the recipes for quasi-static force measurements \cite{backholm2019}.
A glass capillary tube is pulled asymmetrically to achieve a long tapered end. Using a microforge, the tapered micropipette is cut to size and bend twice by 90$^{\circ}$, yielding the typical double-L shape shown in \textbf{Fig.~\ref{Fig1}a}.
Alterations of the pulling speed and variations in the shaping of the cantilever allow to adjust the nozzle size and cantilever stiffness.
The force sensor is immersed in a liquid cell mounted onto an inverted microscope, which is supported by an active vibration-cancelling table.
By attaching a syringe to the pipette and placing it below the level of the liquid cell, i.e.\ by using a hydrostatic pressure difference, a motile \textit{C.~reinhardtii} cell can be partially aspirated at the tip of the micropipette force sensor. 
Both flagella beat freely in the liquid medium during all experiments, see \textbf{Fig.~\ref{Fig1}b}.

\begin{figure}[t]
\centering
	\includegraphics[width=1\columnwidth]{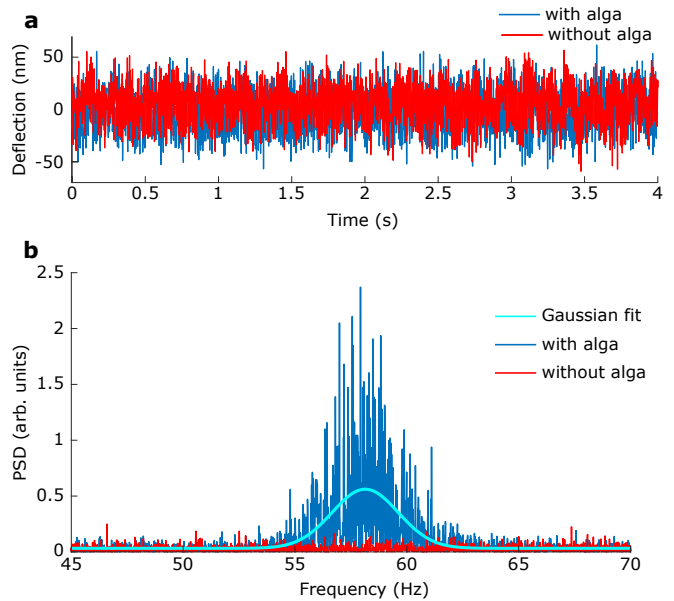}
\caption{Force sensor signal analysis.
\textbf{a} Micropipette deflection with and without a cell attached (overall recording time of the experiment was 82\,sec).
\textbf{b} Power spectra of the experimental deflection data shown in (\textbf{a}). The power spectrum of the deflection with a cell attached (blue) reveals a distinct signal between 55 and 60\,Hz, whereas the power spectrum of the same micropipette without cell attached (red) shows noise. The solid line represents a Gaussian fit to the cell signal.}
\label{Fig2}
\end{figure}

\subsection{Data Acquisition \& Signal Analysis}
\label{Signal Analysis}

We now measure the oscillatory forces and flagella beating frequency of \textit{C.~reinhardtii} cells in bulk conditions.
In order to extract the micropipette deflection during the microbe's swimming motion, we record the micropipette cross-section close to the nozzle with a 40x objective, see \textbf{Fig.\ \ref{Fig1}d} and \textbf{e}. 
A set of data consists of $N=2^{15}$ cross-section images taken at $f_\mathrm{sample}=400$ frames per second (fps), resulting in an observation time of about 82 seconds. 
We determine the micropipette's deflection for any given frame as the shift of the current pipette position relative to the pipette position in the first frame using a cross-correlation analysis of the intensity profiles of the pipette cross-section, see \textbf{Fig.~\ref{Fig1}f}, following established protocols \cite{kreis2018,backholm2019}. 
Determining the micropipette position as precisely as possible is crucial for the accuracy of the dynamic force measurement.
We find that the pixel-wise correlation analysis, where the maximal correlation alternates between two pixels, does not result in a sufficient spatial resolution and introduces an error in the force measurements. 
Interpolating the correlation curve and extracting the shift corresponding the the maximum of the interpolation, however, reveals a fluctuating signal with sub-pixel resolution, see \textbf{Fig.~\ref{Fig1}g}.

While the micropipette deflection with and without a cell attached appear similar in real space, see \textbf{Fig.\ \ref{Fig2}a}, a distinct signal originating from the oscillatory forcing of the flagella can be identified in the frequency domain, see \textbf{Fig.\ \ref{Fig2}b}.
The single-sided power spectrum (normalized squared Fourier coefficients, abbreviated as PSD) of the micropipette fluctuations displays a clear cell signal in the frequency range between 50 to 65\,Hz, which agrees well with the typical flagella beating frequency of \textit{C.~reinhardtii} \cite{Rueffer1985}. 
This signal disappears once the cell is released from the micropipette nozzle.
The Fourier transformation $X(k)$ of an $N$-pointed array $x(n)$ used to calculate the power spectrum is defined as

\begin{align}
X(k+1)=\sum_{n=1}^{N} x(n+1) e^{-2\pi i k n /N}.
\end{align}

Note that despite the fact that all experiments were performed using an active vibration-cancelling table, the full power spectrum may still show signatures of building vibrations at low frequencies and/or electrical devices (e.g.\ camera fans) at high frequencies. 
Since none of these external signals fall into the frequency range associated to the flagella beating, they do not interfere with the signal originating from the cell in Fourier space.
The beating frequency $f$ of the flagella fluctuates over time and the frequency distribution may well be described by a Gaussian distribution \cite{Wan2014}, thus, we use a Gaussian fit to extract the cell signal and isolate it from external noise, see \textbf{Fig.\ \ref{Fig2}b}. 
The fit function of the Gaussian distribution is defined as
\begin{align}
\label{eq:Gauss}
g(f)=a \cdot \exp \left( -\frac{(f-\mu)^2}{2\sigma^2}\right)+d,
\end{align}
with amplitude $a$, mean beating frequency $\mu$, variance $\sigma$ and constant offset $d$. 
By subtracting the offset from the fit, we eliminate the white noise contribution from the power spectrum. 
Based on the Gaussian fit to the experimental data, the signal analysis in the power spectrum allows to quantify the flagella's beating  its variance through the parameters $\mu$ and $\sigma$. Using Parseval's theorem, we can also extract the signal power associated to the oscillatory beating of the flagella $P'_{\text{c}}$ as the area of the Gaussian curve minus the offset: 
\begin{align}
P'_{\text{c}}=\frac{\sqrt{2 \pi} \sigma a}{\Delta f}.
\end{align}
The associated error of $P'_{\text{c}}$ is calculated using error propagation of the uncertainties of the fit parameters $\sigma$ and $a$. 
Note that the power spectrum, along with $P'_{\text{c}}$, is given in units of squared deflection, since we have not yet taken into account the frequency response and spring constant of the micropipette cantilever. 
Also, it is important to distinguish between periodic forcing and mean propulsion force generated by the beating flagella.
Due to the definition of the Fourier transform, $P'_{\text{c}}$ is composed of forces exerted by the beating flagella that can be constructed from superposition of sinusoidal functions. 
Consequently, $P'_{\text{c}}$ only captures the oscillatory forcing of the flagella motion with zero mean forcing. 
Note that the mean propulsive force is ideally a constant offset in the pipette deflection (or a peak at zero frequency in the power spectrum) and cannot be readily accessed with this approach. 

\begin{figure}[t]
\centering
	\includegraphics[width=1\columnwidth]{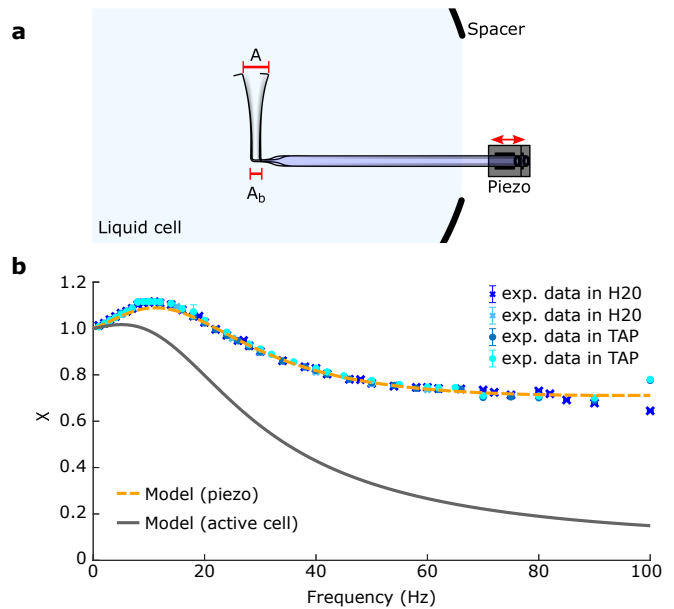}
\caption{Dynamic calibration.
\textbf{a} Setup and relevant parameters for the experimental calibration of the frequency response $\chi=A/A_{\text{b}}$.
\textbf{b} Frequency response $\chi$ of the micropipette force sensor. Experimental data obtained in water and TAP medium (symbols) and the theoretical model (dashed line) fitted to the experimental results validate the applicability of the theory and allows to calculate the frequency response for boundary conditions during measurements with an active cell (solid line).}
\label{Fig3}
\end{figure}

\section{Dynamic Force Calibration}
\label{OM dynamic calibration}

In order to convert the pipette fluctuations into forces, the micropipette force sensor has to be calibrated. 
For static and quasi-static force measurements, i.e.\ at low frequencies of at most a few Hz, it suffices to determine the micropipette's spring constant $k$. 
The static calibration is performed by utilizing the changing weight of an evaporating water droplet attached to the tip of the micropipette \cite{kreis2018,backholm2019}. 
Typical spring constants of the pipettes used in our experiments are around 1\,nN/\textmu{}m.
Performing measurements on the dynamics of cellular appendages at high frequencies requires a full characterization of the micropipette's frequency response.
Determining the frequency response under the exact same conditions found throughout an experiment, i.e.\ with an oscillating forcing applied to the pipette nozzle and a stationary micropipette base, is experimentally challenging. 
We quantify the frequency response under more accessible boundary conditions, where the micropipette is actuated inside the liquid cell with a sinusoidal motion at its clamped body by a closed-loop piezo actuator, as shown in \textbf{Fig.~\ref{Fig3}a}. 
The experimental calibration data can then be translated into the response for the boundary conditions found throughout experiments with an active cell using the theoretical model outlined in Section \ref{OM model}.
In brief, we model the micropipette force sensor as a tapered, hollow elastic beam that is oscillating in a viscous liquid.
The dynamical deflection of the pipette is described with the Kirchhoff equations, coupled to the time-dependent Stokes equations to take viscous and inertial fluid drag into account \cite{Sader1998}. Inertia and hydrodynamic drag of the nozzle are implemented as a localized dynamic loading at the end of the elastic beam. 
In the model, the boundary conditions found throughout the measurements on active cells correspond to a concentrated time-dependent loading at the tip of a cantilever, while the body of the micropipette is stationary. 
The calibration measurements correspond to a free tip and a clamped body that is oscillating in normal direction.

\textbf{Figure~\ref{Fig3}\textbf{b}} shows the frequency response $\chi$ of the micropipette, normalised to the static response. For calibration boundary conditions, i.e.\ actuation with a piezo, experimental (symbols) and theoretical (dashed line) results are in excellent agreement. 
For the boundary conditions of experiments with an active cell we calculate the corresponding response of the micropipette cantilever (\textbf{Fig.~\ref{Fig3}\textbf{b}}, solid line). 
The error of the response curve is estimated by the mean deviation between the experimental and model calibration data and translated to the active cell case as a relative error.

\section{Elasto-Hydrodynamic Model}
\label{OM model}

In order to model the oscillation of the micropipette in a liquid medium, we simplify the pipette geometry, treating the cantilever part as a straight, tapered beam that is oscillating at small amplitude normal to its long axis. 
The nozzle of the pipette is treated as a long, thin rod that is oscillating in longitudinal direction. Inertial and drag forces from the nozzle are then coupled to the cantilever as a concentrated load at the cantilever's end, together with the forces generated by attached cells. 
To model the cantilever motion, we closely follow the approach by Sader \cite{Sader1998}, using the Kirchhoff equation for a tapered beam,
\begin{equation}
\label{eq:kirchhoff}
\frac{\partial^2}{\partial x^2}\left(B(x) \frac{\partial^2 w(x,t)}{\partial x^2}\right) + m(x) \frac{\partial^2 w(x,t)}{\partial t^2}  = F(x,t).
\end{equation}
Here, $x\in [0,L]$ is the coordinate along the cantilever of length $L$ and $w$ is the displacement of the cantilever from its resting position perpendicular to its long axis.
\begin{align}
	B(x) &= \frac{\pi}{4} E \left( R(x)^4 - r(x)^4 \right)\\
	m(x) &= \pi \rho_p \left(R(x)^2 - r(x)^2\right) + \pi \rho\, r(x)^2
\end{align}
are bending modulus and mass per unit length of the tapered cantilever, respectively. 
$E$ and $\rho_p$ are Young's modulus and density of the pipette material, $r$ and $R$ are inner and outer Radius, respectively, and $\rho$ is the density of the medium inside the pipette. 
$F(x,t)$ is the spatio-temporal loading distributed along the cantilever, which is given by viscous and inertial drag of the surrounding fluid medium. 
The spring constant can be derived from the static version of Eq.~(\ref{eq:kirchhoff}), which is obtained by setting $\partial w(x,t)/\partial t \equiv 0$ and $F(x,t) \equiv 0$.

In order to quantify the hydrodynamic drag in the dynamic case, we first analyze the relevant length scales in the problem. These are the amplitude of the oscillation, $A\sim 0.1$\,\textmu m, the outer diameter of the pipette, $2R\sim 50$\,\textmu m, and the thickness of the viscous boundary layer $\lambda = \sqrt{2\eta/\rho \omega}\sim 70$\,\textmu m, where $\eta$ is the viscosity of the medium and $\omega$ is the angular frequency of the oscillation. 
In our case, $A\ll a\sim\lambda$, which is a small amplitude oscillation in a viscous medium. 
We define the Reynolds number for the local acceleration term,
\begin{equation}
Re = \rho \omega R^2/\eta\sim 0.2
\end{equation}
and, accordingly, describe the fluid velocity $u$ and pressure $p$ by the time-dependent Stokes equation~\cite{Williams1972,Sader1998},
\begin{equation}
\label{eq:stokes}
\rho\frac{\partial u}{\partial t} = -\nabla p + \eta\,\nabla^2 u,
\end{equation}
supplemented by the continuity equation for an incompressible fluid, $\nabla\cdot u = 0$, a no-slip boundary condition at the surface of the cantilever, and require a quiescent fluid at large distance. 
Solving the fluid mechanics problem for the actual shape of the cantilever would require full numerical simulations, coupled to Eq.~(\ref{eq:kirchhoff}). 
However, the taper angle of the cantilever is very small and it can locally be approximated by a cylindrical beam, for which analytical solutions to the time dependent Stokes equation exist.

Due to the linearity of Eqs.~(\ref{eq:kirchhoff}) \& (\ref{eq:stokes}), they are best solved after Fourier-transformation with respect to time, yielding
\begin{equation}
\label{eq:model}
\frac{\mathrm{d}^2}{\mathrm{d} x^2}\left(B(x) \frac{\mathrm{d}^2 \hat w(x)}{\mathrm{d} x^2}\right) - m(x) \omega^2 \hat w(x) = \hat F(x)
\end{equation}
and
\begin{equation}
- i\rho\omega \hat u = -\nabla \hat p + \eta\dot\nabla^2 \hat u,
\end{equation}
in which the `` $\hat\;$ '' indicates transformed quantities and the angular frequency $\omega$ merely acts as a parameter~\cite{Sader1998}. The solution of the time-dependent Stokes equation is a lengthy derivation that can be found in~\cite{Tuck1969}. Here, we repeat the resulting net load per unit length onto the cantilever~\cite{Sader1998},
\begin{equation}
\hat F(x) = \pi \rho\,\omega^2 R(x)^2 \Gamma(x,\omega)\,\hat w(x),
\end{equation}
with the hydrodynamic function $\Gamma(\omega)$ that reads, for oscillation perpendicular to the long axis~\cite{Sader1998},
\begin{equation}
\Gamma(x,\omega) = 1 + \frac{4i K_1(-i\sqrt{i\, Re})}{\sqrt{i\, Re} K_0(-i\sqrt{i\, Re})},
\end{equation}
where $K_n$ are the modified Bessel functions of the third kind and $n$-th order. Note that in this formulation, $\Gamma$ and $Re$ have to be evaluated for the local cross-section and, for a tapered beam, thus depend on $x$.

The boundary conditions are an external load $f$ that, together with the hydrodynamic drag of the nozzle, is concentrated at $x=L$. 
We approximate the nozzle as a long, thin cylinder of radius $R(L)$ and length $L_n$, and thus describe it by the solution to the axisymmetric version of Stokes' second problem~\cite{Tuck1969}. The boundary condition at $x=L$ is:
\begin{align}
&-\left. \frac{\partial}{\partial x}\left( B(x)\frac{\partial_x^2 \hat w(x)}{\partial x^2} \right)\right|_{x=L} = \hat f - \nonumber\\
& \quad L_n \left( \pi \rho R(L)^2 \Gamma_n(\omega) + m(L) \right)\omega^2\,\hat w(L),\\
&\hat w''(L) = 0,
\end{align}
where $\Gamma_n(\omega) = (\Gamma^*(L,\omega) - 1)/2$ is the hydrodynamic function for a long cylinder in axial oscillation and ``~$^*$~'' denotes conjugation. At the clamped end of the beam at $x=0$, an oscillation with angular frequency $\omega$ and amplitude $w_0$ is prescribed:
\begin{equation}
\hat{w}(0) = w_0,\quad\quad \hat w'(0)=0.
\end{equation}
In calibration conditions, $\hat f=0$ and $w_0\neq 0$; in measurement conditions, $\hat f\neq 0$ and $w_0=0$. 
With these boundary conditions, Eq.~(\ref{eq:model}) is solved numerically to obtain the spring constant and resonance curves.
Since the resonance curves strongly depend on the pipette geometry, we allow for global additive corrections to the measured geometry parameters, $r(x)+\delta r $, $R(x) + \delta R$, $L+\delta L$, and $L_n + \delta L_n$, that are obtained by fitting the resonance curve to the experimental calibration data. Fitting is performed by a finite differencing scheme of the numerical resonance curves. 
Importantly, the correction values are constrained to be within the uncertainty of the respective measurements. $\delta r$, $\delta R$, and $\delta L$ must fulfil an additional relation that matches the spring constant in the model to the experimentally determined one. 
The resulting correction values are below one micron for the radii and on the order of $10\,$\textmu m for the lengths.
We provide a Mathematica script as \textbf{Supplementary Code} that demonstrates the numerical solution of the equations above to obtain the calibration and measurement resonance curves displayed in \textbf{Fig.~\ref{Fig3}b}.

\begin{figure}[t]
\centering
	\includegraphics[width=.49\textwidth]{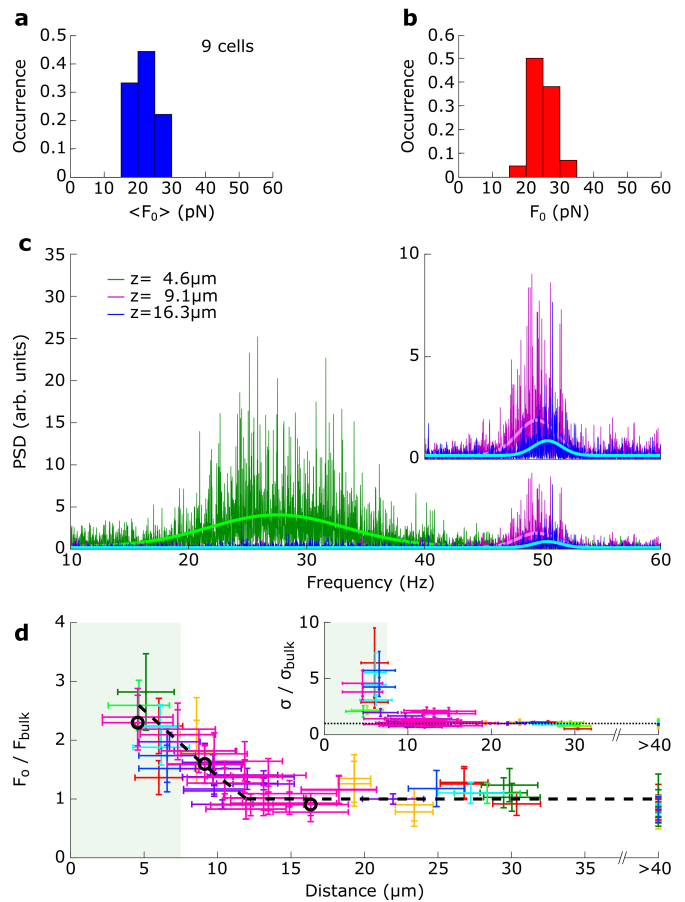}
\caption{Dynamic force measurements on \textit{C.~reinhardtii} cells in bulk liquid and in proximity to a solid/liquid interface.
\textbf{a} Distribution of mean values of the forces $F_{\text{0}}$ for 9 different cells recorded in bulk.
\textbf{b} Distribution of forces $F_{\text{0}}$ from 42 bulk measurements for the same cell.
\textbf{c} PSDs of the same cell at distance $z=4.6\, \mu$m (green, $\mu=27.5$\,Hz, $F_{\text{0}}=60.6\, \pm$ 4.2\,pN), at $z=9.1\,\mu$m (purple, $\mu=49.5$\,Hz, $F_{\text{0}}=42.1\, \pm$ 3.0\,pN) and at $z=16.3\, \mu$m (blue, $\mu=50.4$\,Hz, $F_{\text{0}}=23.9\, \pm$ 1.9\,pN) from the wall.
\textbf{d} Force $F_{\text{0}}$ as a function of distance $z$ from the wall for 8 different cells normalized by their respective bulk values $F_{\text{bulk}}$. 
Data corresponding to different cells are distinguished by different colors. 
The dashed line serves as a guide to the eye; open circles indicate the datasets displayed in \textbf{(c)}.
The area shaded in green between 0 and 7.5\,\textmu{}m corresponds to the regime where the flagella may exhibit steric interactions with the interface, as determined from flagellar beating pattern in bulk conditions shown in \textbf{Fig.~\ref{Fig1}\textbf{b}}.
The inset displays the corresponding normalized variance $\sigma/\sigma_{\text{bulk}}$ of the cell signals in the PSD.
}

\label{Fig5}
\end{figure}

\section{Oscillatory Forcing of Beating Flagella}

The final step of the analysis is the conversion of the signal power $P'_{\text{c}}$, see Section \ref{Signal Analysis}, to physical quantities using the dynamic force calibration outlined above. 
Note that an inverse Fourier transform does not suffice to recover the motion of the flagella. 
We have lost the information about the phase throughout the signal analysis by squaring all coefficients to calculate the power spectrum and by fitting the signal with a Gaussian in order to eliminate noise. 
Instead, we perform a manual inverse Fourier transform by projecting $P'_{\text{c}}$ into a sinusoidal with frequency $\mu$ to recover the periodic forcing $F_{\text{0}}(t)$, using

\begin{align}
\label{eq:force}
F_{\text{0}}(t)= \sqrt{\frac{2 P'_{\text{c}}}{N}} \cdot \frac{k}{\chi(\mu)} \sin(\mu t),
\end{align}
with spring constant $k$ and frequency response $\chi(\mu)$.
The error of $F_{\text{0}}$ is calculated using error propagation of individual errors of $P'_{\text{c}}$, $k$ and $\chi(\mu)$, where the error of $k$ is taken as the standard deviation of all experimental spring constant calibrations for a given micropipette. 

Overall 109 individual measurements with 9 different cells were performed in bulk.
The experiments yielded a mean amplitude of the periodic forcing of 23\,pN with a standard deviation of 5\,pN and a mean beating frequency of 51\,Hz with a standard deviation of 6\,Hz. 
Histograms of the mean dynamic force $F_{\text{0}}$ of 9 different cells (at least two measurements per cell) and of 42 measurements of the same cell recorded over a time period of more than two hours are shown in \textbf{Fig.~\ref{Fig5}a} and \textbf{b}, respectively.
Cell-cell variations and variations of the dynamic forces of a single cell over time are larger than the error of individual force measurements (1.8\,pN on average).

In the following, we extend our experiments towards dynamic force measurements on living \textit{C.~reinhardtii} cells in the vicinity of a solid wall in order to study the effect of flagella/wall interactions and, in particular, characterize the presence and magnitude of hydrodynamic interactions between beating flagella and the surface of the wall.
\textbf{Figure \ref{Fig5}c} displays three representative PSDs for the exact same cell, recorded at different distance $z$ between the cell body and the surface of a silicon wafer.
For $z=4.6\, \mu$m we find a lowered mean beating frequency $\mu$ as compared to the PSD recorded at $z=16.3\, \mu$m, which can be safely considered bulk swimming.
Besides the shift of the mean frequency, also the variance of the frequency $\sigma$ is found to increase substantially, which indicates that steric interactions alter the regular beating pattern.
Interestingly, already at intermediate distances, i.e.\ $z=9.1\, \mu$m, the signature of the flagella beating in the PSD differs significantly from the bulk measurement, as shown in the inset of \textbf{Fig.\ \ref{Fig5}c}.
This is a first indication that hydrodynamic interactions might indeed be detectable in recorded cell signals, given that the flagella of \textit{C.~reinhardtii} only have a maximal forward reach of 7 to 8\,\textmu{}m in the unperturbed beating pattern (see \textbf{Fig.~\ref{Fig1}b}). 

We studied overall 8 different cells at varying distance $z$ between the base of the flagella and the wall and recorded the respective dynamic forces $F_{\text{0}}$ following the approach outlined in the previous sections and Eq.~(\ref{eq:force}).
Note that we normalise $F_{\text{0}}$ of a given cell by the respective value in bulk conditions to account for cell-cell variations.
Based on our results, we consider the experimental data recorded at distances over 15\,\textmu{}m from the wall as bulk measurements. 
The data for $F_{\text{0}}/F_{\text{bulk}}$ are shown in \textbf{Fig.~\ref{Fig5}d}, while the inset displays the corresponding normalized variance $\sigma/\sigma_{\text{bulk}}$ of the flagella beating frequencies in the PSD.
Significant deviations of the forcing from the bulk values occur at distances smaller than about 12\,\textmu{}m from the wall.
Below this threshold we find a monotonic increase of the normalized dynamic force $F_{\text{0}}/F_{\text{bulk}}$ for decreasing distance of the beating flagella to the wall.
The fact that the dynamic forcing is already significantly enhanced at distances greater than the maximal flagella forward reach of 7 to 8\,\textmu{}m, e.g.\ we find $F_{\text{0}}=42.1\, \pm$ 3.0\,pN at $z=9.1\,\mu$m as compared to $F_{\text{0}}=23.9\, \pm$ 1.9\,pN in bulk for the datasets shown in \textbf{Fig.~\ref{Fig5}c}, strongly suggests an effect of hydrodynamic interactions.
In the regime below distances of 7 to 8\,\textmu{}m we expect steric interactions between the flagella and the wall to dominate, in line with the finding that the force transduction is further enhanced ($F_{\text{0}}=60.6\, \pm$ 4.2\,pN for the datasets shown in \textbf{Fig.~\ref{Fig5}c}).
As shown in the inset of \textbf{Fig.~\ref{Fig5}d}, the normalized variance of the beating frequency $\sigma/\sigma_{\text{bulk}}$ abruptly increases at distances of 7 to 8\,\textmu{}m to the wall, which we attribute to flagella desynchronization due to transient flagella-substrate contact during every beating cycle. 
Thus, $\sigma/\sigma_{\text{bulk}}$ as a function of the distance confirms the determination of the maximum forward reach of the flagella from the beating pattern.

\section{Discussion}

We established a direct experimental approach based on dynamic micropipette force sensors (DMFS) for precise and robust measurements of the dynamic forces generated by periodically beating cellular appendages at high frequencies.
The observed mean beating frequency of \textit{C.~reinhardtii} of $51\pm6$\,Hz falls into the expected range of beating frequencies from 40 to 64\,Hz~\cite{Rueffer1985} and matches very well the previously reported mean beating frequency of $53\pm5$\,Hz from high-speed cell tracking~\cite{Guasto2010}. 
Our \textit{in vivo} force measurements on \textit{C.~reinhardtii} yielded a mean amplitude of the periodic forcing of $23\pm5$\,pN provided by the coordinated beating of both flagella.
Assuming a sinusoidal functional form of the oscillatory forcing and adding the mean propulsion force as a constant offset, we can construct a minimal model of the instantaneous forcing throughout one beating cycle.
By extracting the swimming velocity from cell tracks and applying Stokes drag the mean propulsion force of \textit{C.~reinhardtii} has been estimated to be $8\pm 2$\,pN \cite{Minoura1995}, which results in a maximum forcing of $31\pm 7$\,pN.
This value is in quantitative agreement with the measured escape forces of 26 to 31\,pN of \textit{C.~reinhardtii} in optical tweezer experiments~\cite{McCord2005}. 
Assuming sinusoidal forcing, our dynamic force measurements yield that 39\,\% [31 44] (square brackets indicate lowest and highest error bounds) of the beating cycle show negative instantaneous forces, i.e.\ correspond to the recovery stroke of the beating flagella, which is in accordance with previous calculations of the mean instantaneous cell body speed~\cite{Guasto2010} as well as hydrodynamic simulations~\cite{Geyer2013}.

In conclusion, we provided the first direct measurements of the dynamic forces of swimming microorganisms originating from the beating of their flagella, in varying environmental conditions.
We established our experimental approach for \textit{C.~reinhardtii} cells in bulk liquid swimming conditions. 
With this approach we measured the dynamic forces exerted by a pair of beating flagella in the vicinity of a solid surface.

The natural habitats of many microorganisms, including soil-dwelling \textit{C.~reinhardtii} microalgae, are confined spaces.
Consequently, their navigation and motility is governed by frequent interactions with the confining walls.
By characterizing the nature of the interactions of flagellated puller-type microbes encountering surfaces quantitatively, we are able to gain new insights regarding the microbial motility in complex natural and physiological environments.
We find a gradual increase of the dynamic force transduction as we decrease the flagella-wall distance, well before steric contact interactions might take over, and attribute this regime to the presence of hydrodynamic interactions. 
This finding is relevant far beyond the microbial domain since eukaryotic flagella are well-preserved phenotypical features that exist in virtually all eukaryotic life forms, including the human body.

On the methodological side, we developed a novel experimental approach for direct dynamic force measurements with pN resolution, extending the applicability of micropipette force sensors from quasi-static to dynamic force measurements.
The provided calibration protocol for determining the full frequency response function of the force sensor ultimately enables these quantitative dynamic force measurements. 
Under controlled experimental conditions the DMFS method allows for extracting periodic forces from the power spectrum of the micropipette deflection data, after isolating the signal from background noise. 
The force measurements are based on the optical readout of the micropipette deflection with sub-pixel resolution, yielding a force sensitivity in the range of a few piconewton. 

The experimental technique is highly versatile and allows to vary physiologically relevant parameters, such as illumination, exposure to drugs, external flow, and proximity to interfaces, even during on-going measurements. 
In combination with the ability to perform repeated measurements per cell over extended periods of time, DMFS appears as a unique tool to systematically quantify the effect of these parameters on microbial propulsion and the activity of cellular appendages. 
In the future, the full optical access to the biological sample may allow for simultaneously combining dynamic force measurements with flagella tracking, PIV measurements as well as contrast-enhanced optical techniques to observe the biological processes at work. 
Since DMFS can be readily implemented into existing experimental setups and by using tailored micropipette force sensors with customiszed geometry and sensitivity, we expect that in the future DMFS can be employed to measure dynamic forces of a variety of living systems, including microorganisms of different size and shape.


\section{Acknowledgements}
The authors are grateful to K.\ Dalnoki-Veress, K.\ Wan, B.\ Friedrich, H.\ Nobach and M.\ Lorenz for discussions. M.M.\ Makowski is acknowledged for technical assistance. The G\"ottingen Algae Culture Collection (SAG) kindly provided the \textit{C.~reinhardtii} wild-type strain SAG 11-32b.\\

O.B.\ conceived the project. T.J.B.\ and O.B.\ designed the experiments. T.J.B.\ performed the experiments and analyzed the data. T.J.B.\ and S.K.\ developed the dynamical model. T.J.B., C.T.K., Q.M.\ and O.B.\ developed the measurement protocol. T.J.B.\ wrote the first draft of the paper. All authors contributed to the preparation of the manuscript.\\



\section{Appendix:\ Materials \& Methods}

\subsection{Cell cultivation}
\label{OM cell cultivation}
All experiments have been performed with wild-type \textit{Chlamydomonas reinhardtii} strain SAG 11-32b. 
Axenic cell cultures were grown in tris-acetate-phosphate (TAP) medium (reference number A13798-01, Thermo Fisher Scientific, USA) on a 12:12\,h day-night cycle in a Memmert IPP 100Plus incubator (Memmert, Germany).
The daytime temperature was set to 24$^{\circ}$C with a light intensity of 1 to $2\cdot 10^{20}$ photons/m$^2$s. 
During the night cycle, the light intensity was zero and the temperature was reduced to 22$^{\circ}$C.
Experiments have been performed during the daytime with vegetative cells in the logarithmic growth phase on the second to fourth day after incubation. 
Cells were suspended in a liquid cell filled with TAP under ambient conditions (24 -- 26$^{\circ}$C).

\begin{figure}[b]
\centering
	\includegraphics[width=.49\textwidth]{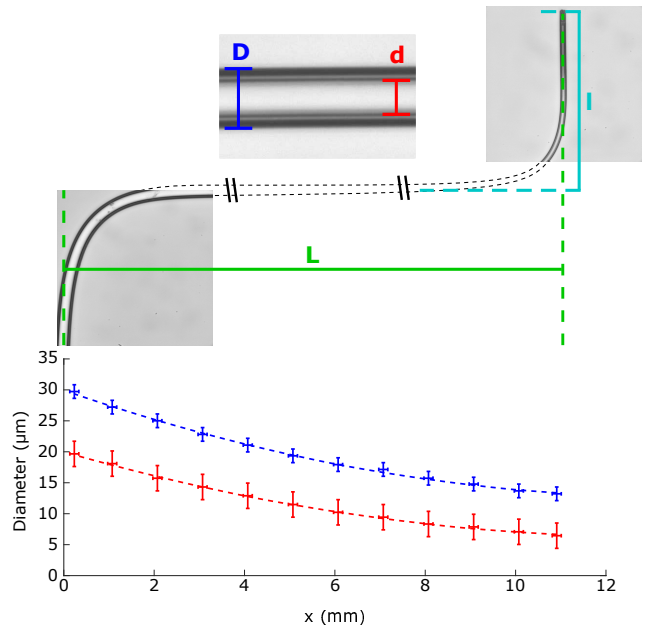}
\caption{Precise measurement of the micropipette's geometric parameters. Inner and outer diameter of the micropipette cantilever $d$ and $D$ were measured based on optical micrographs of the micropipette (typical error of 3\,\textmu{}m) taken at equidistant intervals along the micropipette cantilever. 
The cantilever length $L$ was measured based on the movement of the micromanipulator; the nozzle length $l$ was directly obtained from micrographs. 
To extract the precise shape of the micropipette, we fit a second order polynomial to the acquired data for the inner and outer diameter.
}
\label{appendix1}
\end{figure}

\subsection{Micropipette fabrication \& calibration consistency checks}
\label{OM Micropipette fabrication}
The micropipette cantilevers were custom-made and adjusted in size and stiffness. 
A blank glass capillary (TW100-6, World Precision Instruments, USA, outer diameter 1\,mm, inner diameter 0.75\,mm) was pulled to the desired diameter using a micropipette puller (P-97, Sutter Instruments, USA). During the pipette pulling process, one side of the capillary was held stationary, while the other side was slowly pulled away. This yielded one short and one long, tapered pipette.
Using a microforge (MF-900, Narishige Group, Japan), the long pipette was cut to the desired length and bend twice to achieve the desired shape. 
The bend pipette was finally filled with deionised water and placed in a water bath for storage.
To modulate the suction pressure during experiments, the pipette was furnished with a water-filled tube attached to a syringe. 
The cantilever arm had a typical length of about 1\,cm with a final bend leading to the nozzle with a length of about 100 to 200\,\textmu{}m. 
The base of the cantilever arm had an outer diameter typically ranging between 20 to 30\,\textmu{}m, tapering to an outer diameter of less than 15\,\textmu{}m at the nozzle. 
The inner diameter at the nozzle tip was approximately 5\,\textmu{}m.
As shown in \textbf{Fig.\ \ref{appendix1}}, the geometric dimensions of the micropipette force sensors were extracted from micrographs taken along the cantilever arm and nozzle. 
Subsequently, the measurements of inner and outer diameter were fitted with a second-order polynomial.

\begin{figure}[t]
\centering
	\includegraphics[width=.49\textwidth]{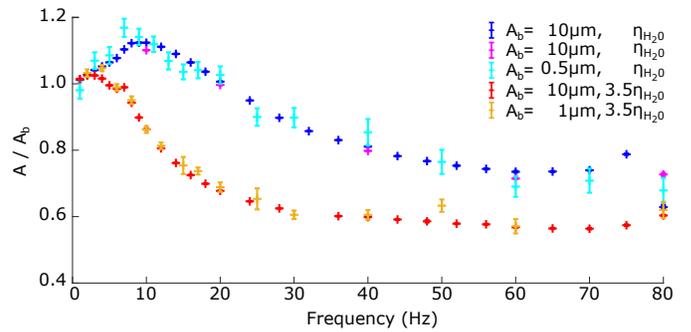}
\caption{Validation of the experimental dynamic calibration method to measure the micropipette force sensor's frequency response $\chi=A/A_{\text{b}}$.
Variation of the amplitude of the external actuation of the pipette base $A_{\text{b}}$ did not show any significant effect on the frequency response of the cantilever. 
Upon increasing the viscosity of the surrounding liquid (using a water-glycerol mixture), the damping of the pipette increases as expected for increased hydrodynamic drag.
}
\label{appendix2}
\end{figure}

We performed consistency checks to verify the experimental method used to determine the frequency response of the micropipette force sensor. 
\textbf{Figure \ref{appendix2}} shows the frequency response of the force sensor under different amplitude of the external forcing $A_{\text{b}}$. 
The frequency response $\chi=A/A_{\text{b}}$ was found to be independent of the amplitude within the range of applied amplitudes of up to 10\,\textmu{}m, which is about two orders larger than the pipette deflection found in the experiments with cells.
To quantify the effect of the viscosity $\eta$ of the liquid surrounding the pipette on the frequency response, we performed the calibration in water and in a mixture of water and glycerol with approximately 3.5 times the viscosity of water. 
As expected for increased hydrodynamic drag, the damping increased in the later case. 
The difference in viscosity of distilled water and TAP growth medium, however, was found to have a negligible effect on the micropipette's frequency response, as shown in \textbf{Fig.\ \ref{Fig3}b}.

\subsection{Experimental procedure}
\label{OM Experimental procedure}
Experiments were performed in a liquid cell consisting of two glass slides (\textsc{2947-75x50, Corning, USA}) held in place by custom-made frames. Two spacers cut from an o-ring of 3\,mm thickness and an inner diameter of 3.8\,cm defined the experimental chamber. 
The glass slides were cleaned with ethanol (Uvasol, Merck, Germany) and completely dried before assembling the cell. 
The o-rings were also cleaned with ethanol, rinsed with water and then dried. 
The assembled cell was filled with TAP growth medium and mounted on an inverted microscope (IX83, Olympus, Japan). 
Liquid \textit{C.\ reinhardtii} culture was injected into the cell. 
The culture was not centrifuged before injection but taken directly out of the liquid cultures in the incubator. 
The amount of cell culture injected into the liquid cell was adjusted such that a sufficient amount of \textit{C.~reinhardtii} were found in the chamber.

A thin slab of silicon cut from a wafer (type P/Bor, orientation $\braket{100}$, resistivity 1–20\,\si{\ohm}cm, unilateral polished, Si-Mat, Germany) has been glued onto a custom-made stainless steel substrate holder with PDMS. 
Substrate and holder were prepared for experiments by sonication in an ethanol bath and subsequent blow-drying with an air gun. 
The substrate was mounted onto a linear stage (M462 series, Newport Corporation, USA) capable of computer-controlled movement within the focal plane and mechanical adjustment of the height and pitch. 
The substrate was then inserted into the liquid cell and moved to the focal point of the microscope. 
Due to the phototaxis of \textit{C.~reinhardtii}, the algae can be guided towards the substrate in the light beam focused on the substrate. 
Under white light conditions, the flagella stick to surfaces such that the algae accumulate on the substrate \cite{kreis2018,kreis2019}.
The micropipette cantilever, furnished with a tube and syringe filled with deionised water, was mounted onto a 3-axis piezo-driven manual micromanipulator (PCS-5400, Burleigh, Thorlabs, USA) and inserted into the liquid cell. 
The substrate was then adjusted such that the surface was orthogonal to the pipette nozzle.
An individual cell adhering to the substrate was approached by the pipette and fixated at the pipette nozzle with a suction pressure adjusted manually with the syringe. 
The fixated alga could then be detached from the substrate by retracting the pipette. 
Since only the flagella stuck to the substrate \cite{kreis2018}, algae pulled off the substrate were oriented such that the flagella faced away from the pipette. 
Thus, the direction of the propulsion forces was orthogonal to the cantilever arm and the flagella were beating freely. Note that the flagella were not damaged when pulled off the substrate and recovered their regular beating pattern within tens of seconds \cite{kreis2018}.
For measurements in bulk conditions, the substrate was typically retracted for more than one millimetre from the pipette.   
In red light, the flagella-mediated adhesion of \textit{C.~reinhardtii} to substrates is inhibited \cite{kreis2018}, thus enabling to probe the steric interactions between flagella and substrate.
Measurements during which the flagella, even just transiently, attached to the substrate were discarded.
In order to ensure comparability between bulk measurements and measurements in the proximity of the interface, a 550\,nm cut-off filter was inserted into the light path for all experiments.
The aperture of the microscope was completely opened during experiments and the light intensity was kept constant to ensure consistent illumination.
The entire setup was placed on an active anti-vibration table (Halcyonics i4-large, Accurion, Germany) and contained in a closed box to minimize acoustic vibrations and to control the illumination conditions during the experiment.

One data set consisted of 2$^{15}$ monochromatic images of the cross-section of the micropipette cantilever recorded close to the end of the cantilever arm using a 40x objective and a high-speed camera at 400\,fps (pco.edge 4.2, PCO, Germany). 
The relative shift of the micropipette position throughout the recording was determined using a cross-correlation analysis of the intensity profiles of the pipette cross-section, following established protocols \cite{kreis2018,backholm2019}. 
The intensity profile was extracted from the center line of each image. The profile of the first frame served as a reference for the correlation analysis against which the profiles of all subsequent images were correlated. 
Subtracting the position of the maximal correlation value in the auto-correlation of the reference profile from the position of the maximum of the cross-correlation of the reference profile with the intensity profiles in subsequent images gives the relative shift of the pipette position compared to the first frame.
Interpolating the correlation curve using the Matlab function 'smoothingspline' over 40 pixel around the discrete maximal correlation value and extracting the shift corresponding the the maximum of the interpolation results in sub-pixel resolution.

\subsection{Analysis of the power spectrum and noise}
\label{OM Spectrum and noise}

\begin{figure}[b]
\centering
	\includegraphics[width=.49\textwidth]{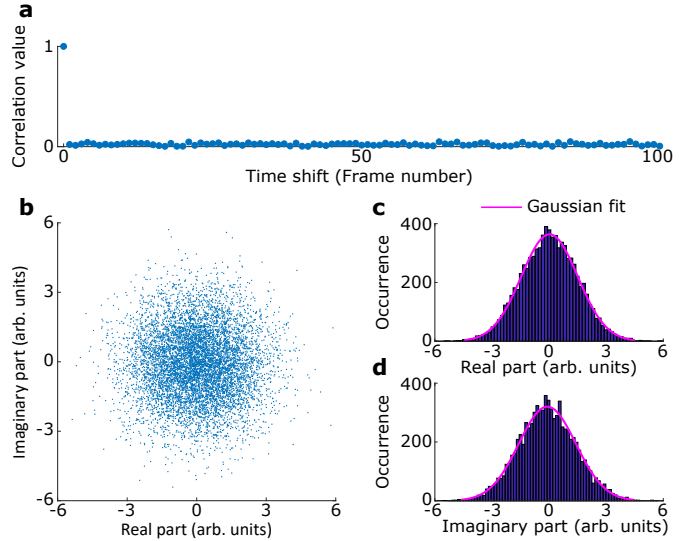}
\caption{Noise analysis of the power spectra obtained without a cell attached to the force sensor.
\textbf{a} Correlation of the noise in the power spectrum shown in \textbf{Fig.\ \ref{Fig2}} in the range of 30 to 130\,Hz calculated from the inverse Fourier transform of the power spectrum following the Wiener-Khinchin Theorem. The noise is found to be uncorrelated in time.
\textbf{b} Fourier coefficients extracted from the same power spectrum as shown in \textbf{(a)}. 
\textbf{c} Histogram of the real part of the scatter plot shown in \textbf{(c)}. The Gaussian fit yields a mean of $0.023\,\pm 0.033$ with a variance of $1.482\,\pm0.023$.
\textbf{d} Histogram of the imaginary part of the scatter plot shown in \textbf{(c)}. The Gaussian fit yields a mean of $-0.092\,\pm 0.034$ with a variance of $1.501\,\pm0.023$.
}
\label{appendix3}
\end{figure}

Following the Wiener-Khinchin theorem, we calculated the correlation of the noise in the range of 30 to 130\,Hz from the power spectrum without a cell attached to the micropipette (shown in \textbf{Fig.\ \ref{Fig2}}) using the inverse Fourier transform. 
As sown in \textbf{Fig.\ \ref{appendix3}a}, the correlation revealed a delta peak for zero shift only, i.e.\ the noise is uncorrelated in time as expected for white noise.
\textbf{Figure \ref{appendix3}b} displays the scatter of the Fourier coefficients of the same data. 
The Fourier coefficients were found to follow a Gaussian distribution for both real (see \textbf{Fig.\ \ref{appendix3}c}) and imaginary (see \textbf{Fig.\ \ref{appendix3}d}) part centred around zero with finite variance. 
From this analysis, we conclude that this is white noise.

Assuming that the micropipette is a linear system, as is the Fourier transform, the white noise in the power spectrum can be accounted for with the offset $d$ of the Gaussian fit as illustrated in Eq.\,(\ref{eq:Gauss}):
Due to linearity, each Fourier coefficient, represented as a signal vector, $\vec{s}$, in complex space, is constructed as the linear combination of an signal vector originating from the forcing of the cell, $\vec{c}$, and a noise signal vector $\vec{n}$. 
Following the law of cosines, the amplitude of the resulting signal vector, $|\vec{s}|=s$, can be calculated from the cell signal amplitude, $c$, and noise signal amplitude, $n$, as
\begin{align}
s^2=c^2+n^2-2 cn \cos(\alpha).
\end{align}
The phase $\alpha$ takes a random value since it is the combination of the phase of $\vec{c}$ and the phase of $\vec{n}$, the latter being random as discussed above.
Averaging over multiple realisations yields
\begin{align}
<s^2>=<c^2> +<n^2>-2<c  n \cos(\alpha)>,
\end{align}
where `` $< >$ '' indicates averaged quantities.
Since the probability distribution of $\cos(\alpha)$ is symmetric around zero, the last term cancels out. 
If the frequency sampling is sufficiently high, i.e.\ the frequency binning-width $\Delta f=f_{\text{sample}}/N$ is small such that there is a large number of coefficients within a frequency interval over which the fitted frequency distribution changes significantly, fitting a distribution to the power spectrum is the same as locally averaging over many realisations.
For a typical experiment in bulk conditions, the standard deviation of the Gaussian fit $\sigma$ varied between 1 to 3\,Hz, which was considerably larger than the frequency binning-width $\Delta f=0.012$\,Hz.
Consequently, white noise was treated as a constant offset in the power spectrum, which can be subtracted in order to isolate the signal originating from the periodic forcing of the attached cell.



\end{document}